\documentclass[a4paper, 10pt]{extarticle}
\usepackage[a4paper, top=2cm, bottom=2cm, left=2cm, right=2cm, marginparwidth=1.75cm]{geometry} 
\usepackage[english]{babel}
\usepackage[utf8]{inputenc}
\usepackage[T1]{fontenc}
\usepackage{float}
\usepackage{authblk}
\usepackage{amsmath}
\usepackage{amssymb}
\usepackage{amsthm}
\usepackage{mathrsfs}
\usepackage{comment}
\usepackage[font=small]{caption}
\usepackage{tabularx}
\usepackage{graphicx}
\usepackage{marvosym}
\usepackage{pst-node}
\usepackage{arydshln}
\usepackage{multirow}
\usepackage{tikz-cd}
\usepackage{float}
\usepackage{algorithm2e}
\usepackage{xcolor}
\usepackage[colorlinks=true, allcolors=blue]{hyperref}
\definecolor{blue}{rgb}{0,.45,1}
\definecolor{red}{rgb}{1,.3,0}
 
\RestyleAlgo{boxruled}

\begin{document}

\title{Supervised Kernel PCA For Longitudinal Data}
\author{Patrick Staples}
\author{Min Ouyang}
\author{Robert F. Dougherty}
\author{Gregory A. Ryslik}
\author{Paul Dagum}
\affil{Mindstrong Health}

\maketitle

\begin{abstract}

In statistical learning, high covariate dimensionality poses challenges for robust prediction and inference. To address this challenge, supervised dimension reduction is often performed, where dependence on the outcome is maximized for a selected covariate subspace with smaller dimensionality. Prevalent dimension reduction techniques assume data are \emph{i.i.d.}, which is not appropriate for longitudinal data comprising multiple subjects with repeated measurements over time. In this paper, we derive a decomposition of the Hilbert-Schmidt Independence Criterion as a supervised loss function for longitudinal data, enabling dimension reduction between and within clusters separately, and propose a dimensionality-reduction technique, \emph{sklPCA}, that performs this decomposed dimension reduction. We also show that this technique yields superior model accuracy compared to the model it extends.
\end{abstract}

\pagebreak

\section{Introduction}
High covariate (feature) dimensionality poses challenges for robust prediction and inference, and dimensionality reduction is often performed to address this challenge. Supervised dimensionality reduction techniques \cite{bair2006prediction} attempt to preserve the dependence of the outcome on a reduced covariate subspace while minimizing the dimensionality of that subspace. Supervised kernel dimension reduction (\emph{SKDR}) methods provide increased flexibility in specifying the relationship between covariates and outcomes by determining the distance metric between points in an implied (often high-dimensional) Hilbert space rather than the dependence of covariates and outcomes directly. Gretton \emph{et al.} \cite{gretton2005measuring} proposed an independence criterion in reproducing kernel Hilbert spaces \cite{scholkopf2002learning}, the Hilbert-Schmidt Independence Criterion (\emph{HSIC}). Many proposed covariate selection methods use penalized feature weighting to optimize dependence via the \emph{HSIC} \cite{allen2013automatic} \cite{cao2007feature} \cite{fukumizu2012gradient} \cite{grandvalet2003adaptive} \cite{sun2014kernel} \cite{weston2003use}. A stepwise selection procedure was suggested by Song \emph{et al.} \cite{song2012feature}. The approach by Chen \emph{et al.} \cite{chen2017kernel} seeks to find a covariate subset for which the \emph{HSIC} is minimized conditional on that set. Other methods yield a selection of feature transformations, such as Barshan \emph{et al.} \cite{barshan2011supervised}, who propose a method that selects a subset of feature eigenvectors.

Each of these methods assume data are \emph{i.i.d.} (independent and identically distributed), which is not true for clustered data. Such data arise naturally in the setting of longitudinal data, which consists of repeated measurements on multiple subjects. Mixed model approaches such as generalized estimating equations \cite{hardin2002generalized} and generalized linear mixed models \cite{mcculloch2001generalized} maximize the dependency of features on the outcome while accounting for clustering, and penalized methods such as GLMM LASSO \cite{groll2014variable} enable feature selection in this context. The degree of clustering might be of interest, in which case \emph{SKDR} methods that maximize clustering have been developed \cite{song2007dependence}. In the longitudinal setting, clustering may be considered a nuisance characteristic of the data, where the primary interest is maximizing the dependence between features and the outcome within clusters. To the authors' knowledge, no literature to date proposes \emph{SKDR} techniques in the context of longitudinal data.

This paper derives a method for \emph{SKDR} in the longitudinal data setting, extending the methods proposed by Barshan \emph{et al} \cite{barshan2011supervised}. In Section 2, we review the definition of the \emph{HSIC} and propose a decomposition of the \emph{HSIC} resulting in fixed and random components. We derive estimators for this decomposition and prove their bias and rate of convergence. We also propose an algorithm for maximizing these estimators while selecting a feature subspace with reduced dimension. In Section 3, we perform a simulation study to evaluate the properties of this method and compare this derivation with existing methods, and apply this method to the Michigan Intern Health Study. We close with final remarks in Section 4. 

\section{Methods}

\subsection{Definition and Decomposition of the \emph{HSIC}}

Consider random variables $X$  and $Y$ representing $p$ distinct features and univariate outcomes, respectively. Let $\mathcal{F}$ and $\mathcal{G}$ be separable Hilbert spaces, equipped with reproducing kernels $k: X\times X\rightarrow \mathbb{R}$ and $l: Y\times Y\rightarrow \mathbb{R}$ such that $k(x, x') = \langle\varphi(x), \varphi(x')\rangle$ and $\ell(y, y') = \langle\psi(y), \psi(y')\rangle$ for some maps $\varphi:X\rightarrow \mathcal{F}$ and $\psi:Y\rightarrow \mathcal{G}$, where $\langle\cdot, \cdot \rangle$ denotes an inner product. Gretton \emph{et al.}\cite{gretton2005measuring} suggest a measurement of dependence in terms of kernels for the features and outcomes:

\begin{align}
\textit{HSIC}(\mathbb{P}_{X, Y}, \mathcal{F}, \mathcal{G}) :&=\|\mathbb{E}_{X, Y}[(\varphi(x)-\mu_{\varphi(X)})\otimes(\psi(y)-\mu_{\psi(Y)})]\|^2_{\text{HS}} \label{HSIC}
\end{align}

where $\otimes$ is the tensor product, $\mu_{\varphi(X)}=\mathbb{E}_X(\varphi(x))$ and $\mu_{\psi(Y)}=\mathbb{E}_Y(\psi(y))$. This may also be formulated in terms of kernel expectations:

\begin{align}
\textit{HSIC}(\mathbb{P}_{X, Y},\mathcal{F}, \mathcal{G}) &= \mathbb{E}_{X,X',Y,Y'}[k(x,x')\ell(y,y')] \nonumber \\
&\hspace{1cm}-2\mathbb{E}_{X,Y}[\mathbb{E}_{X'}[k(x,x')]\mathbb{E}_{Y'}[\ell(y,y')]] \nonumber \\
&\hspace{1cm}+\mathbb{E}_{X,X'}[k(x,x')]\mathbb{E}_{Y,Y'}[\ell(y,y')] 
\end{align}

Random variables $X$ and $Y$ are independent when this value is zero. For \emph{i.i.d.} data, \emph{SKDR} consists of minimizing $\text{dim}(\xi(X))$ for some selection or transformation map $\xi$ while maximizing $\textit{HSIC}(\mathbb{P}_{\xi(X), Y},\mathcal{F}, \mathcal{G})$.

If a collection of covariate sets is available and a transformation of reduced dimension is desired for each set, Theorem 1 provides a formulation of the joint \emph{HSIC} in terms of these random variables.

\textbf{Theorem 1}: \emph{Let $X_1$ and $X_2$ be random variables, $\mathcal{F}_1$ and $\mathcal{F}_2$ be separable Hilbert spaces equipped with kernels $k_1: X_1\times X_1\rightarrow \mathbb{R}$ and $k_2: X_2\times X_2\rightarrow \mathbb{R}$ such that $k_1(x_1, x_1') = \langle\varphi(x_1)$, $\varphi(x_1')\rangle_{\mathcal{F}_1}$ and $k_2(x_2, x_2') = \langle\varphi(x_2), \varphi(x_2')\rangle_{\mathcal{F}_2}$ for some feature maps $\varphi_1: X_1 \rightarrow \mathcal{F}_1$ and $\varphi_2: X_2\rightarrow \mathcal{F}_2$. Let $X_1\oplus X_2$ be the direct sum of these random variables, and $\mathcal{F}_1\oplus \mathcal{F}_2$ be the direct sum of these Hilbert spaces $\mathcal{F}_1$ and $\mathcal{F}_2$. Also let $Y$, $\mathcal{G}$, $\ell$, and $\psi$ be defined as before. Then:} 

\begin{align}
\emph{HSIC}(\mathbb{P}_{X_1\oplus X_2, Y}, \mathcal{F}_1\oplus \mathcal{F}_2, \mathcal{G}) =  \emph{HSIC}(\mathbb{P}_{X_1, Y}, \mathcal{F}_1, \mathcal{G}) + \emph{HSIC}(\mathbb{P}_{X_2, Y}, \mathcal{F}_2, \mathcal{G}).
\end{align}

Proof for Theorem 1 is provided in the Supplementary Materials. This theorem readily generalizes to any finite collection of random variables.  In general, the random variables and kernels used for between- and within-subject terms are arbitrary. Examples of separate covariate sets suitable for these terms would be subject-level demographic information and within-subject repeated measures. In the context of mixed models, model fits are sought in terms of between- and within-group covariates. Consider a general distribution $Z$, representing cluster membership. Theorem 2 provides a decomposition for \emph{HSIC} for a single covariate set using a generalization of the law of total covariance.

\textbf{Theorem 2}: \emph{Let $X$, $Y$, and $Z$ be random variables, $\mathcal{F}$ and $\mathcal{G}$ be separable Hilbert spaces equipped with kernels $k: X\times X\rightarrow \mathbb{R}$ and $l: Y\times Y\rightarrow \mathbb{R}$ such that $k(x, x') = \langle\varphi(x), \varphi(x')\rangle_{\mathcal{F}}$ and $\ell(y, y') = \langle\psi(y), \psi(y')\rangle_{\mathcal{G}}$ for some explicit feature maps $\varphi:X\rightarrow \mathcal{F}$ and $\psi:Y\rightarrow \mathcal{G}$. Then:}

\begin{align}
\emph{HSIC}(\mathbb{P}_{X, Y}, \mathcal{F}, \mathcal{G}) &= \underbrace{\left\|\text{Cov}_Z(\mu_{\varphi(X\mid z)}, \mu_{\psi(Y\mid z)})\right\|_{\text{HS}}^2}_{\emph{HSIC}_{\text{fixed}}} + \underbrace{\mathbb{E}_Z\left\|\text{Cov}_{X, Y \mid z}(\varphi(x), \psi(y))\right\|_{\text{HS}}^2}_{\emph{HSIC}_{\text{random}}} + \zeta(\mathbb{P}_{X, Y, Z}, \mathcal{F}, \mathcal{G}). \label{conditional_HSIC}
\end{align}

Theorem 2 is proven in the Supplementary Materials. In this equation, $\zeta(\mathbb{P}_{X, Y, Z}, \mathcal{F}, \mathcal{G})$ is a nuisance cross-covariance term. In the context of mixed models, the first two terms represent the dependence of the outcome on fixed effects and random effects, respectively. Estimates of these terms may then be used in turn to reduce the dimensions of the fixed effect and random effect covariates.

The estimands of interest in Theorem 2 can also be cast in terms of Theorem 1, by treating these estimands as separate subspaces. Theorem 1 assumes that the total \emph{HSIC} term is a direct sum of random variables and Hilbert spaces, and any covariance between these terms is assumed to be zero. In mixed regression models, fixed and random effects are typically modeled as independent of each other; we therefore restrict our attention to maximizing the first two terms in Theorem 2.  In this paper, we consider the decomposition of a single covariate set only.

\subsection{Estimators of \emph{HSIC}}

With our estimands defined, we now derive estimators of these quantities. Let subjects $i=1,...,m$ have repeated measurements $j=1,..., n_i$, comprising $n:=\sum_{i=1}^m n_i$ observations in total. For Subject $i$, let observation $j$ with $p$ features be written $\mathbf{x}_{ij}:=[x_{ij1}, ..., x_{ijp}]$, let $\mathbf{X}_i:=[\mathbf{x}_{i1}^\top, ..., \mathbf{x}_{in_i}^\top]^\top$ be a $n_i\times p$ matrix of feature data for Subject $i$, and let feature data for all subjects be written $\mathbf{X}_{n\times p}:=[\mathbf{X}_1^\top, ..., \mathbf{X}_m^\top]^\top$. Similarly, let $\mathbf{Y}_i:=[y_1,...,y_{n_i}]^\top$ and $\mathbf{Y}:=[\mathbf{Y}_1^\top,...,\mathbf{Y}_m^\top]^\top$ represent the outcome data. Let $\mathbf{K}$ and $\mathbf{L}$ be the sample kernels for all observations, where $\mathbf{K}_{i,i'} = k(\mathbf{X}_{i,\cdot}, \mathbf{X}_{i', \cdot})$ and $\mathbf{L}_{i,i'} = \ell(\mathbf{Y}_i, \mathbf{Y}_{i'})$, respectively. Finally, let $\mathbf{1}_m := [1,...,1]^\top_{m\times 1}$, and $\mathbf{I}_m:=\text{diag}(\boldsymbol{1}_m)$, and let the $m\times m$ centering matrix be defined as $\mathbf{H}_m:=\mathbf{I}_m-m^{-1}\mathbf{1}_m\mathbf{1}_m^\top$.

\textbf{Fixed Effects}

Let selection matrix $\mathbf{S}_i$ be defined as:

\begin{align}
\mathbf{S}_i &:= 
\begin{bmatrix}
\mathbf{0}_{(\sum_{i'<i}n_{i'})\times n_i} \\
\mathbf{I}_{n_i} \\
\mathbf{0}_{(\sum_{i<i'}n_{i'})\times n_i} \\
\end{bmatrix}
\end{align}

For matrix $\mathbf{W}_{n\times n}$, $\mathbf{S}_i^\top\mathbf{W}$ and $\mathbf{W}\mathbf{S}_i$ result in the selection of the rows and columns associated with Subject $i$, respectively. Let $\overline{k}_{i,i'} := \mathbb{E}_{X\mid z_i, X'\mid z_{i'}}\left[k(x,x')\right]$ and $\overline{\ell}_{i,i'} := \mathbb{E}_{Y\mid z_i, Y'\mid z_{i'}}\left[\ell(y, y')\right]$. We define the kernel sample mean between two subjects $i$ and $i'$ as follows:

\begin{align}
\overline{\mathbf{K}}_{i,i'} &:= (n_i-1)^{-1}(n_{i'}-1)^{-1}\mathbf{1}_{n_i}^\top\mathbf{S}_i^\top\mathbf{K}\mathbf{S}_{i'}\mathbf{1}_{n_{i'}} \\
\overline{\mathbf{L}}_{i,i'} &:= (n_i-1)^{-1}(n_{i'}-1)^{-1}\mathbf{1}_{n_i}^\top\mathbf{S}_i^\top\mathbf{L}\mathbf{S}_{i'}\mathbf{1}_{n_{i'}}
\end{align}

Let $\overline{\mathbf{K}}_m:=[\overline{\mathbf{K}}_{i,i'}]_{m\times m}$ and $\overline{\mathbf{L}}_m:=[\overline{\mathbf{L}}_{i,i'}]_{m\times m}$. We propose the following estimator for $\emph{HSIC}_{\text{fixed}}$:

\textbf{Theorem 3}: \emph{Under regulatory conditions, then $\widehat{\text{HSIC}}_{\emph{fixed}} := (m-1)^{-2}\emph{tr}(\overline{\mathbf{K}}_m\mathbf{H}_m\overline{\mathbf{L}}_m\mathbf{H}_m)$ is a consistent estimator of $\text{HSIC}_{\emph{fixed}}$ with bias of order $\mathcal{O}(m^{-1})$ and convergence rate $\mathcal{O}(m^{-1/2})$.}

Proof for Theorem 3 is provided in the Supplementary Materials.

\subsubsection{Random Effects}

For random effects, we must specify the within-subject correlation structure, analogous to choosing a covariance structure in mixed estimation \cite{zhang1998semiparametric}. For example, conditional independence holds if within-subject observations are independent, a strong but common assumption. The following estimator assumes conditional independence holds. Let $\mathbf{K}_i := \mathbf{S}_i^\top\mathbf{K}\mathbf{S}_i$ and $\mathbf{L}_i := \mathbf{S}_i^\top\mathbf{L}\mathbf{S}_i$. A simple unbiased estimator for the expected covariance across subjects is the sample average of subject estimates:

\begin{equation}
\widehat{\emph{HSIC}}_{\text{random}} = m^{-1}\sum_{i=1}^m(n_i-1)^{-2}\text{tr}\left(\mathbf{K}_i\mathbf{H}_{n_i}\mathbf{L}_i\mathbf{H}_{n_i}\right)
\label{HSIC_CI_form} \\
\end{equation}

Theorems 1 and 3 in \cite{gretton2005measuring} imply that the summands in $\widehat{\emph{HSIC}}_{\text{random}}$ each exhibit bias of order $\mathcal{O}(n_i^{-1})$ and convergence with rate $\mathcal{O}(n_i^{-1/2})$. In terms of Theorem 1, we can combine both \emph{HSIC} terms into a single estimator by treating them as the direct sum of their respective spaces:

\begin{align}
\widehat{\emph{HSIC}}_{\text{mixed}} &:= \widehat{\emph{HSIC}}_{\text{fixed}} + \widehat{\emph{HSIC}}_{\text{random}} \\
&= \text{tr}((m-1)^{-2}\overline{\mathbf{K}}_m\mathbf{H}_{m}\overline{\mathbf{L}}_m\mathbf{H}_{m}) + m^{-1}\sum_{i=1}^{m}\ (n_i-1)^{-2}\text{tr}(\mathbf{K}_i\mathbf{H}_{n_i}\mathbf{L}_i\mathbf{H}_{n_i}) \label{HSIC_spaces} \\
&= \text{tr}\{(m-1)^{-2}\overline{\mathbf{K}}_m\mathbf{H}_{m}\overline{\mathbf{L}}_m\mathbf{H}_{m}\ \underset{i}{\bigoplus}\ m^{-1}(n_i-1)^{-2}\mathbf{K}_i\mathbf{H}_{n_i}\mathbf{L}_i\mathbf{H}_{n_i}\}
\end{align}

Because each of these terms is non-negative and independent, the global maximum of $\widehat{\emph{HSIC}}_{\text{mixed}}$ is the maximum of each component.

\subsection{Maximization of \emph{HSIC} Estimators}

To perform \emph{SKDR} in the $i.i.d.$ case, Section 5.3.2 of Barshan \emph{et al.}\cite{barshan2011supervised} proposes selecting the top generalized eigenvectors of $(\mathbf{K}\mathbf{H}_n\mathbf{L}\mathbf{H}_n\mathbf{K}, \mathbf{K})$. We denote this approach as supervised kernel PCA, or \emph{skPCA}. We now proceed with a similar derivation for closed-form solutions for maximizing $\widehat{\emph{HSIC}}_{\text{mixed}}$ while reducing the dimension of components for use in longitudinal mixed models. Let $\xi_{\text{fixed}}(\mathbf{X})$ and $\xi_i(\mathbf{X})$ be to fixed and subject-specific maps of $\mathbf{X}$ with reduced dimension. We seek $\xi_{\text{fixed}}(\mathbf{X}), \{\xi_i(\mathbf{X})\}_n$ that maximize $\widehat{\emph{HSIC}}_{\text{mixed}}$ while minimizing $\text{dim}\large(\xi_{\text{fixed}}(\mathbf{X})\large)$, $\text{dim}\large(\{\xi_i(\mathbf{X})\}_n\large)$.

\subsubsection{Fixed Effects}

Consider feature map $\overline{\varphi}$ that yields the subject-level mean of $\varphi(\mathbf{X})$ such that $\overline{\mathbf{K}}_m=\overline{\varphi}(\mathbf{X})\overline{\varphi}(\mathbf{X})^\top$. Let $\mathbf{U}$ be a operator from $\overline{\varphi}$ to the subspace of $\overline{\varphi}$ spanned by $\overline{\varphi}(\mathbf{X})$. Then $\overline{\mathbf{K}}_m$ can be rewritten as $\overline{\varphi}(\mathbf{X})\mathbf{U}\mathbf{U}^\top\overline{\varphi}(\mathbf{X})^\top$. Using the Representer Theorem \cite{kimeldorf1971some}, let $\mathbf{U}$ be written as a linear combination $\overline{\varphi}(\mathbf{X})^\top \overline{\mathbf{V}}_m$. Then $\overline{\mathbf{K}}_m=\overline{\varphi}(\mathbf{X})\overline{\varphi}(\mathbf{X})^\top\overline{\mathbf{V}}_m\overline{\mathbf{V}}_m^\top\overline{\varphi}(\mathbf{X})\overline{\varphi}(\mathbf{X})^\top$, and the fixed component estimator can be written as follows:

\begin{align}
\widehat{\emph{HSIC}}_{\text{fixed}} &= (m-1)^{-2}\text{tr}\left(\overline{\mathbf{K}}_m\mathbf{H}_m\overline{\mathbf{L}}_m\mathbf{H}_m\right) \\
&= (m-1)^{-2}\text{tr}(\overline{\varphi}(\mathbf{X})\overline{\varphi}(\mathbf{X})^\top\overline{\mathbf{V}}_m\overline{\mathbf{V}}_m^\top\overline{\varphi}(\mathbf{X})\overline{\varphi}(\mathbf{X})^\top\mathbf{H}_m\overline{\mathbf{L}}_m\mathbf{H}_m) \\
&= (m-1)^{-2}\text{tr}\left(\overline{\mathbf{V}}_m^\top\overline{\varphi}(\mathbf{X})\overline{\varphi}(\mathbf{X})^\top\mathbf{H}_m\overline{\mathbf{L}}_m\mathbf{H}_m\overline{\varphi}(\mathbf{X})\overline{\varphi}(\mathbf{X})^\top\overline{\mathbf{V}}_m\right) \\
&= (m-1)^{-2}\text{tr}(\overline{\mathbf{V}}_m^\top\underbrace{\overline{\mathbf{K}}_m\mathbf{H}_m\overline{\mathbf{L}}_m\mathbf{H}_m\overline{\mathbf{K}}_m}_{\overline{\mathbf{Q}}_m}\overline{\mathbf{V}}_m) \label{HSIC_fixed_final}
\end{align}

subject to $\overline{\mathbf{V}}_m^\top\overline{\mathbf{K}}_m\overline{\mathbf{V}}_m=\mathbf{I}_m$. Solving for $\overline{\mathbf{V}}_m$ is characterized as the generalized eigenvector problem for matrices ($\overline{\mathbf{Q}}_m$, $\overline{\mathbf{K}}_m$). Selecting a subset of these generalized eigenvectors with the largest eigenvalues and computing the kernel composed from training and test features, we may compute the resulting feature subspace that maximizes $\widehat{\emph{HSIC}}_{\text{fixed}}$ for any number of reduced dimensions.

\subsubsection{Random Effects}

Maximizing $\widehat{\emph{HSIC}}_{\text{random}}$ is performed in a similar manner. Let $\mathbf{V}_i$ be similarly defined for all $i$:

\begin{align}
\widehat{\emph{HSIC}}_{\text{random}} &= m^{-1}\sum_{i=1}^m(n_i-1)^{-2}\text{tr}\left(\mathbf{K}_i\mathbf{H}_{n_i}\mathbf{L}_i\mathbf{H}_{n_i}\right) \\
&= m^{-1}\sum_{i=1}^m(n_i-1)^{-2}\text{tr}((\varphi(\mathbf{X}_i)\varphi(\mathbf{X}_i)^\top\mathbf{V}_i\mathbf{V}_i^\top\varphi(\mathbf{X}_i)^\top\varphi(\mathbf{X}_i)\mathbf{H}_{n_i}\mathbf{L}_i\mathbf{H}_{n_i}) \\
&= m^{-1}\sum_{i=1}^m(n_i-1)^{-2}\text{tr}\left(\mathbf{V}_i^\top\varphi(\mathbf{X}_i)\varphi(\mathbf{X}_i)^\top\mathbf{H}_{n_i}\mathbf{L}_i\mathbf{H}_{n_i}\varphi(\mathbf{X}_i)\varphi(\mathbf{X}_i)^\top\mathbf{V}_i\right) \\
&= m^{-1}\sum_{i=1}^m(n_i-1)^{-2}\text{tr}(\mathbf{V}_i^\top\underbrace{\mathbf{K}_i\mathbf{H}_{n_i}\mathbf{L}_i\mathbf{H}_{n_i}\mathbf{K}_i}_{\mathbf{Q}_i}\mathbf{V}_i) \label{HSIC_random_final}
\end{align}

subject to $\mathbf{V}_i^\top\mathbf{K}_i\mathbf{V}_i=\mathbf{I}_m\ \forall i$. Because each term in $\widehat{\emph{HSIC}}_{\text{mixed}}$ is positive, it is maximized when its individual summands are maximized. These steps comprise \emph{sklPCA}, summarized in Algorithm \ref{algorithm_1}.

\begin{algorithm}[H]
\SetAlgoLined
\textbf{INPUT:} Training features $\mathbf{X}_{n\times p}$, training outcomes $\mathbf{Y}_{n\times 1}$, kernels $k$ and $l$, desired dimensionality $q$ and $q_{i^*}$, and test data $\mathbf{X}_{\text{test}}$ with dimension $n_{i^*}\times p$ for Subject $i^*$. \\[.1cm]
\textbf{OUTPUT:} Reduced feature data $\xi_{\text{fixed}}(\mathbf{X}_{\text{test}}), \{\xi_{i^*}(\mathbf{X}_{\text{test}})\}_n$. \\
\begin{enumerate}
\item[1] Compute $\overline{\mathbf{K}}_{i,i'}$ and $\overline{\mathbf{L}}_{i,i'}\ \forall i,i'=1,...,m$.\vspace{-.2cm}
\item[2] Compute kernels $\overline{\mathbf{K}}_{m}$, $\overline{\mathbf{L}}_{m}$, $\mathbf{K}_{i}$, $\mathbf{L}_{i}\ \forall i=1,...,m$.\vspace{-.2cm}
\item[3] Compute $\overline{\mathbf{Q}}_m := \overline{\mathbf{K}}_m\mathbf{H}_m\overline{\mathbf{L}}_m\mathbf{H}_m\overline{\mathbf{K}}_m$ and $\mathbf{Q}_i := \mathbf{K}_i\mathbf{H}_{n_i}\mathbf{L}_i\mathbf{H}_{n_i}\mathbf{K}_i\ \forall i$.\vspace{-.2cm}
\item[4a] Compute $\overline{\mathbf{V}}_m$ := top $q$ generalized eigenvectors from $(\overline{\mathbf{Q}}_m, \overline{\mathbf{K}}_m)$.\vspace{-.2cm}
\item[4b] Compute $\mathbf{V}_i$ := top $q_i$ generalized eigenvectors from $(\mathbf{Q}_i, \mathbf{K}_i)\ \forall i$.\vspace{-.2cm}
\item[5] Compute random test kernel $\mathbf{K}_{i^*} := k(\mathbf{X}_{i^*}, \mathbf{X}_i)\ \forall i=1,...,m$.\vspace{-.2cm}
\item[6a] Compute fixed test kernel $\overline{\mathbf{K}}_{m^*} := [(n_i-1)^{-1}(n_{i^*}-1)^{-1}\mathbf{1}_{n_i}^\top\mathbf{K}_{i^*}\mathbf{1}_{n_{i^*}}]_m$.\vspace{-.2cm}
\item[6b] Compute $\xi_{\text{fixed}}(\mathbf{X}_{\text{test}})$ := $\overline{\mathbf{K}}_{m^*}\overline{\mathbf{V}}_m$.\vspace{-.2cm}
\item[7] $\exists\ i\ (i^*=i)$:\vspace{-.2cm}
\item[] \quad\quad\quad Compute $\xi_{i^*}(\mathbf{X}_{\text{test}})$ := $\mathbf{K}_{i^*}\mathbf{V}_i$. \vspace{-.2cm}
\end{enumerate}
 \caption{\emph{sklPCA}: Perform supervised kernel dimension reduction for longitudinal data.}
 \label{algorithm_1}
\end{algorithm}

\section{Application}
In this section, we demonstrate the utility of \emph{sklPCA} as a supervised kernel dimension reduction method in a longitudinal setting by applying \emph{sklPCA} to simulated and empirical datasets, and compare its performance to \emph{skPCA}.
\subsection{Simulation Specification}

We construct a simulation that satisfies the following conditions: (1) features and outcomes are measured multiple times for several subjects exhibiting within-subject clustering, (2) the relationship between features within and between subjects and the outcome is arbitrary, and (3) features are sampled from a high-dimensional, low-rank space. The modeling task in this simulation is to reduce the dimensionality of these high-dimensional features resulting in between-subject and within-subject components appropriate for use in a mixed model. To elucidate circumstances under which \emph{sklPCA} and/or \emph{skPCA} perform well in this task, we repeat the simulation study under a variety of simulated settings.

For subjects $i=1,...,m$ at measurement time $j=1,...,n_i$, the outcome of interest $\widetilde{\mathbf{Y}}_{ij}$ depends on a subspace $\widetilde{\mathbf{X}}_{ij}$ though an arbitrary function $f_Y$ in a way that differs between and within subjects. The observed outcomes $\mathbf{Y}_{ij}$ are assumed to be noisy measurements of $\widetilde{\mathbf{Y}}_{ij}$, and the observed features $\mathbf{X}_{ij}$ are related to $\widetilde{\mathbf{X}}_{ij}$ linearly through a projection matrix $\mathbf{P}$ with rank $R$ and dimension $D$, where $R\ll D$. The distributional form of these variables are specified as follows:

\begin{align}
\mathbf{Y}_{ij} &= \underbrace{f_Y(\widetilde{\mathbf{X}}_{ij}, \boldsymbol{\mu}_i; \sigma_w) - f_Y(\boldsymbol{\mu}_i, \boldsymbol{0}_R; \sigma_b)}_{\widetilde{\mathbf{Y}}_{ij}} + \epsilon_{ij},\quad \epsilon_{ij} \sim \mathcal{N}(0, \sigma^2_\epsilon)\\
\hspace{2cm}\boldsymbol{\mu}_i &\sim f_\mu(\mathbf{0}_k, \sigma_b^2\mathbb{I}_R),\quad i=1,...,m \\
\widetilde{\mathbf{X}}_{ij} &\sim f_X(\boldsymbol{\mu}_i, \sigma_w^2\mathbb{I}_R),\quad j=1,...,n_i \\
\mathbf{X}_{ij}&=\mathbf{P}^\top\widetilde{\mathbf{X}}_{ij},\quad \mathbf{P} \sim \mathcal{N}(0, 1)^{R\times D}
\end{align}

\subsection{Model Comparison}

We vary four binary factors independently: (1) $\sigma_b/\sigma_w$, or the ratio of between-subject variance compared to within-subject variance, (2) two configurations of distributions applied in the above framework, (3) the rank of the predictors $R$, and (4) the dimension of the predictors $D$. The two distribution configurations are selected to imitate linear and non-linear relationships, respectively:

\begin{enumerate}
\item[]\textbf{Linear Configuration}: $f_\mu, f_X$ are uniform, where the support is defined on $\prod_{r=1}^R [\boldsymbol{\mu}_r\pm\sqrt{3}\sigma]$ to preserve mean $\boldsymbol{\mu}$ and variance $\sigma^2\mathbb{I}_R$. $f_Y^{(\text{linear})}(\mathbf{x}, \boldsymbol{\mu}; \sigma) = \sum_{r=1}^R (\mathbf{x}_r-\boldsymbol{\mu}_r)$, and kernels $k$ and $l$ are set to the linear kernel $\kappa(\mathbf{\mathbf{w}, \mathbf{w}'}; \sigma)=\mathbf{w}^\top\mathbf{w}'$.
\item[]\textbf{Radial Configuration}:  $f_\mu, f_X$ are standard normal, and $f_Y$, $k$, and $l$ are set to the Gaussian radial kernel, $\kappa(\mathbf{w}, \mathbf{w}'; \sigma) = \exp\left(-\frac{1}{2\sigma^2}\left\|(\mathbf{w}-\mathbf{w}')\right\|^2_2\right)$.
\end{enumerate}

\begin{figure}[H]
\begin{center}
\includegraphics[width=15cm]{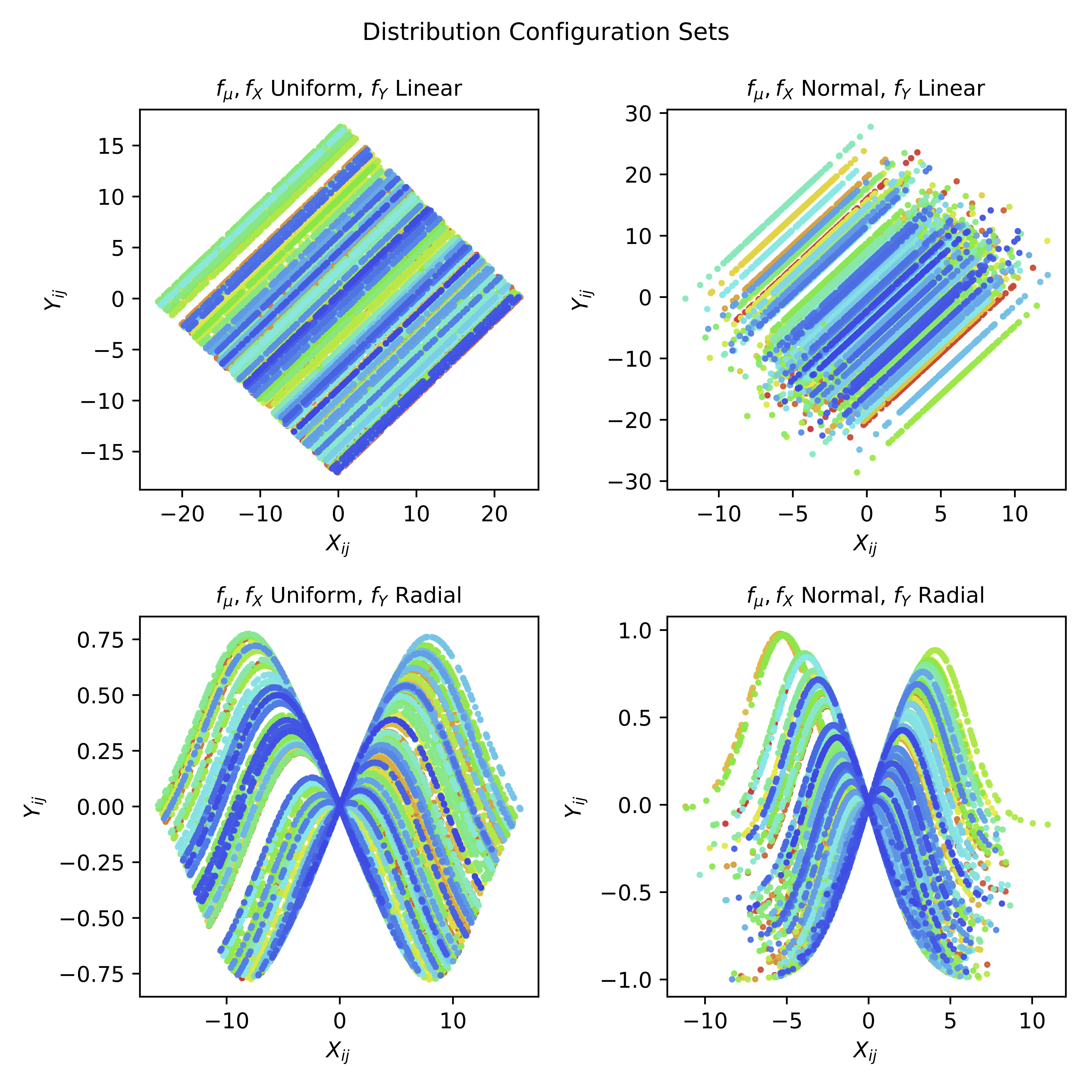}
\caption{Sample datasets using the simulation framework described in this section. The lefthand panels assume $f_\mu, f_X$ are distributed uniformly, whereas the righthand panels assume $f_\mu, f_X$ are distributed normally. The top panels assume $f_Y$ is linear, and the bottom panels assume $f_Y$ takes the form of the Gaussian radial basis function. Individual colors represent separate subjects, and individual points represent feature/outcome measurement pairs. In this figure, $R=D=1$ and $m=n_i=150$ are assumed.}
\label{figure:distribution_configurations}
\end{center}
\end{figure}

Variants of these distribution configurations are displayed visually in Figure \ref{figure:distribution_configurations}; the above configurations correspond to the diagonal panels. Note that $f_Y=f_Y^{(\text{linear})}$ specifies a high-dimensional linear mixed model, whereas $f_Y=f_Y^{(\text{radial})}$ specifies a more complex relationship for which more flexible methods such as \emph{SKDR} is required. The selection of the remaining parameters are $\sigma_b/\sigma_w\in\{0.1, 1\}$, $R\in\{1, 5\}$, and $D\in\{10, 1000\}$. For each combination, several other parameters are held constant. To enable arbitrarily accurate model fits when the model is appropriately specified, the error between the outcome and latent space is assumed to be very low, $\sigma^2_\epsilon=1e-5$. The data are assumed to exhibit moderate size, $m=n_i=50$; for simplicity, we assume $n_i$ is the same for all subjects.

\vspace{-1cm}

After simulation and dimension reduction, linear and mixed regression models are fit to the data reduced using \emph{skPCA} and \emph{sklPCA}, respectively. The mixed models are fit using a two-step estimation approach similar to the unrestricted estimation of longitudinal models \cite{verbeke1997linear}, which is performed by first reducing and fitting the fixed component, then reducing and fitting the random components to the resulting residuals. The criterion of 5-fold cross-validated correlation between the predictions and outcome is used to compare results across each simulation and method. Each parameter combination is simulated 100 times, and the sample mean and standard deviation of the out-of-sample prediction correlations is calculated. Simulations were performed on a 15-inch 2017 MacBook Pro laptop computer, with a 2.8 GHz Intel Core i7 processor (4 cores) and 16 GB RAM. The results of these analyses are shown in Table \ref{table:simulation}.

The general form of $\widetilde{Y}_{ij}$ is designed to produce no little or no correlation when $\sigma_b=\sigma_w$ if subject identity is not taken into account, because the relationship between subjects is the negative of that within each subject. This results in poor model performance when using \emph{skPCA}, which treats each observation independently, whereas \emph{sklPCA} reduces the feature dimension separately based on these separate relationships. When the between-subject variance is small ($\sigma_b/\sigma_w=0.1$), \emph{skPCA} performs reasonably well, although \emph{sklPCA} still yields superior accuracy. Due to the increased complexity, the radial configuration is less performant than the linear configuration. Because the dataset size is held constant, increasing the rank of the feature space has the intuitive effect of reducing overall predictive performance. In contrast, the effect of increasing the dimensionality of the feature space differs for the two distribution configurations: model performance is maintained or slightly improves in the linear case, but model performance decreases in the radial case, and the size of the effect is monotone in $R$ and $D$. For moderate-sized data, determining complex non-linear relationships is particularly challenging in high-rank, high-dimensional settings.

\begin{table}[H]
\begin{center}
\begin{tabular}{ll | rr : rr | rr : rr | l}
 & \multicolumn{1}{c}{} & \multicolumn{4}{c}{$\text{Linear Configuration}$} &\multicolumn{4}{c}{$\text{Radial Configuration}$}\\
 & \multicolumn{1}{c}{} & \multicolumn{2}{c}{$\sigma_b/\sigma_w=0.1$} & \multicolumn{2}{c}{$\sigma_b/\sigma_w=1$} & \multicolumn{2}{c}{$\sigma_b/\sigma_w=0.1$} & \multicolumn{2}{c}{$\sigma_b/\sigma_w=1$} \\ \cline{3-10}
\multirow{4}{*}{$R=1$} & \multirow{2}{*}{$D=10$} &
                                                      0.979 & (0.003) &                   0.050 & (0.050) &                0.601 & (0.051)  &             0.056 & (0.049)              & \multicolumn{1}{l}{\emph{skPCA}} \\
 &&                                                \textbf{0.999} & ($<$1e-3) &   \textbf{0.971} & (0.004) &   \textbf{0.885} & (0.011)  &  \textbf{0.765} & (0.034)  & \multicolumn{1}{l}{\textbf{\emph{sklPCA}}}\\ \cdashline{3-10}
 & \multirow{2}{*}{$D=1000$} &   0.980 & (0.002) &                     0.046 & (0.037) &              0.595 & (0.056)  &              0.044 & (0.047)             & \multicolumn{1}{l}{\emph{skPCA}}\\
 &&                                                \textbf{0.999} & ($<$1e-3) &   \textbf{0.972} & (0.003) &   \textbf{0.785} & (0.024)  &  \textbf{0.631} & (0.047)  & \multicolumn{1}{l}{\textbf{\emph{sklPCA}}}\\ \cline{3-10}
\multirow{4}{*}{$R=5$} & \multirow{2}{*}{$D=10$} &
                                                      0.861 & (0.082) &                   0.124 & (0.047) &               0.674 & (0.046)  &              0.245 & (0.043)              & \multicolumn{1}{l}{\emph{skPCA}}\\
 &&                                                \textbf{0.991} & (0.005) &       \textbf{0.905} & (0.035) &  \textbf{0.813} & (0.027)  &  \textbf{0.667} & (0.030)  & \multicolumn{1}{l}{\textbf{\emph{sklPCA}}}\\ \cdashline{3-10}
 & \multirow{2}{*}{$D=1000$} &   0.964 & (0.017) &                   0.135 & (0.062) &               0.629 & (0.033)  &              0.006 & (0.025)              & \multicolumn{1}{l}{\emph{skPCA}}\\
 &&                                               \textbf{0.998} & ($<$1e-3) &   \textbf{0.958} & (0.005) &   \textbf{0.573} & (0.048)  &  \textbf{0.021} & (0.014)  & \multicolumn{1}{l}{\textbf{\emph{sklPCA}}}\\ \cline{3-10}
\end{tabular}
\caption{Simulation results. Cells represent simulation results varying $\sigma_b/\sigma_w$, distribution configuration, rank $R$, and dimension $D$ independently, comparing \emph{skPCA} and \emph{sklPCA}, where the latter is denoted in bold. Within each cell, mean out-of-sample prediction correlations across 100 repeated simulations is shown on the left (with standard deviations shown in parentheses on the right).}
\label{table:simulation}
\end{center}
\end{table}
\vspace{-.5cm}

\begin{figure}[H]
\begin{center}
\includegraphics[width=15cm]{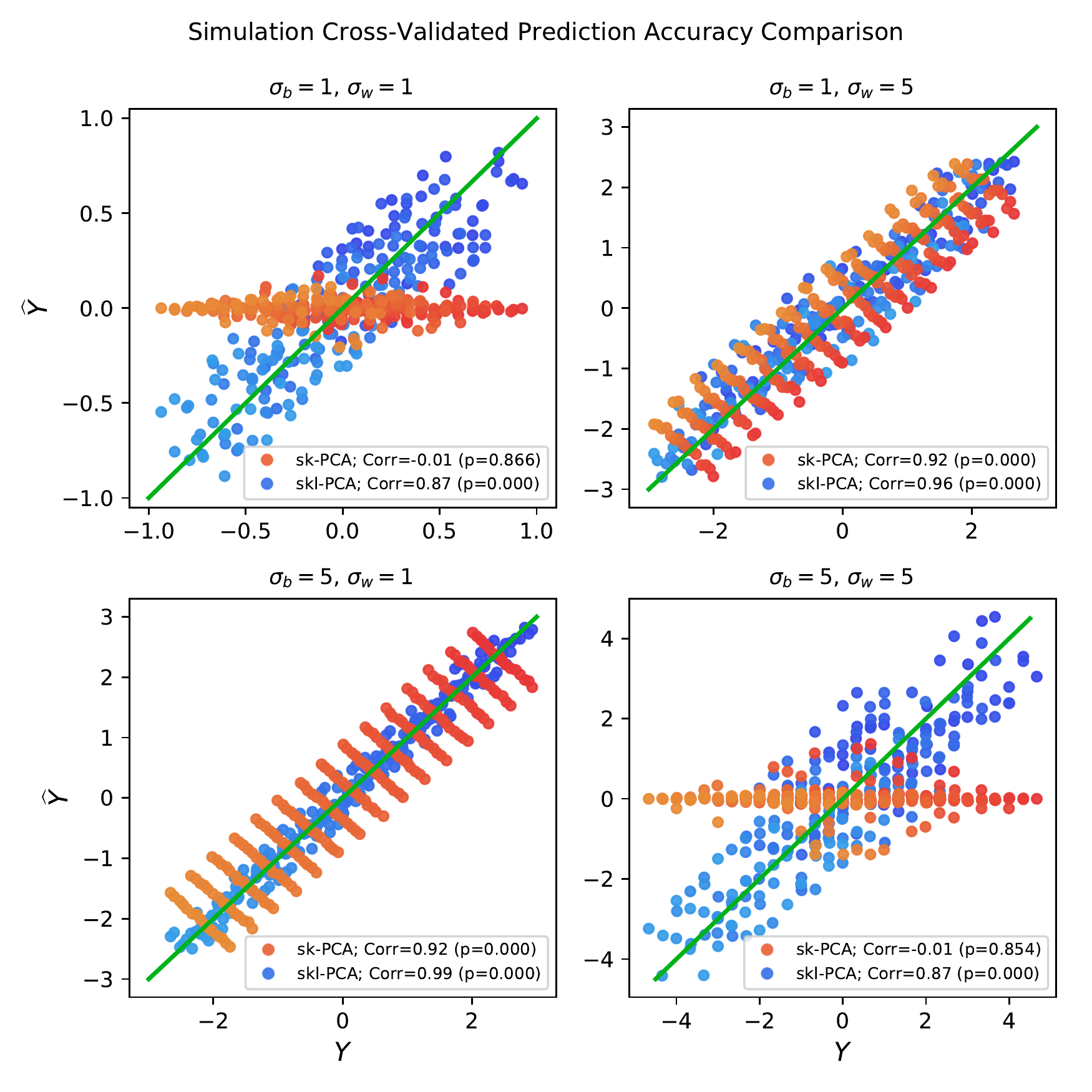}
\caption{A comparison of model performance for the simulated lattice dataset. The top and bottom panels exhibit a between-subject variance of $\sigma_b=1$ and $5$, respectively. Panels on the left side and right side exhibit a within-subject variance of $\sigma_w=1$ and $5$, respectively. In each panel, true simulated outcomes $Y$ are shown on the $x$-axis, and model fits $\widehat{Y}$ are shown on the $y$-axis. The green line is the identity; proximity to this line indicates accurate out-of-sample model performance. The red-like dots depict individual out-of-sample \emph{skPCA} predictions, and shades from red to orange represent different subjects. The blue-like dots depict individual cross-validation \emph{sklPCA} predictions. In each case, the cross-validated correlation using \emph{sklPCA} is stronger than \emph{skPCA}.}
\label{figure:simulation_prediction_accuracy_comparison}
\end{center}
\end{figure}

To illustrate performance differences between these methods, we performed additional simulations designed for visual clarity (see Figure \ref{figure:simulation_prediction_accuracy_comparison}). The relationships between $\widetilde{\mathbf{X}}_{ij}$ and $\widetilde{\mathbf{Y}}_{ij}$ are constructed to produce a regular lattice: $\mu_i = \sigma_b\left(\frac{i}{m}-\frac{1}{2}\right)$, $\widetilde{X}_{ij} = \sigma_w\left(\frac{j}{n_i}-\frac{1}{2}\right) + \mu_i$, and $\widetilde{Y}_{ij} = \sigma_w\left(\frac{j}{n_i}-\frac{1}{2}\right) - \mu_i$. The simulation parameters for this simulation are set as $m=n_i=15$, $\sigma_b \in \{1, 5\}$, $\sigma_w \in \{1, 5\}$, and $\sigma_\epsilon=0.01$. In this figure, perfectly accurate model predictions for all outcomes would yield a diagonal line. For the diagonal panels, the within-subject and between-subject correlations are equal; again, by design, performing data reduction and modeling assuming \emph{i.i.d} data yields little or no model accuracy in expectation. In contrast, the \emph{sklPCA} and mixed model approach does capture the within-subject and between-subject correlations separately, resulting in highly accurate cross-validated predictions. The off-diagonal panels show cases where the variance within subjects and between subjects differ markedly. In these cases, the larger variance component will account for an overall correlation between $\widetilde{X}$ and $Y$, even for the \emph{i.i.d} reduction and modeling strategy. However, Figure \ref{figure:simulation_prediction_accuracy_comparison} shows that predictions using this approach reliably fail to model the remaining correlation resulting from the smaller source of variation; the two sources exhibit correlations of opposite magnitude, and the resulting model fits reveal residuals of potential correlation consistently not captured by this model. In these cases, the \emph{sklPCA} combined with a mixed model approach successfully models both components, yielding comparatively stronger overall correlations.

\vspace{-1cm}

\subsection{Application to the Michigan Intern Health Study}

We now turn to an application of \emph{sklPCA} to an empirical dataset. The Intern Health Study is a multi-site prospective cohort study following physicians throughout internship, for whom correlations are sought between their internship tenure and cognitive and/or clinical outcomes such as mood \cite{kalmbach2018}, depression \cite{sen2010}, and suicidal ideation \cite{guille2015}. Residencies among interns range from internal medicine, general surgery, pediatrics, obstetrics/gynecology, and psychiatry.

This cohort has participated in an assortment of experimental interventions and clinical measurements of well-being. Among these, a subset of physicians installed the Mindstrong digital phenotyping app, which unobtrusively collects content-free feature information such as swiping and typing patterns. A variety of feature extraction techniques are subsequently used to reduce the behavioral gestures to a single, high-dimensional set of features per day per subject. This daily measurement is matched with an outcome of interest: the cumulative sum of minutes of sleep obtained within the last 4 days. The measure of accuracy used to measure model performance is out-of-sample cross-validated correlation, where $k=5$ training folds are created by dividing all paired outcome/feature repeated measurements per subject into a partition of $k$ time-contiguous parts each, for which all but one part are used as training data, and the remaining part is used for predictive testing.

Among the 181 consented physicians, 108 installed the Mindstrong app and uploaded sufficient phone log data from which to extract features and reported minutes spent sleeping per day. We perform a complete cases analysis, in which a subset of the data exhibiting no missing values is selected, comprising 575 features and 56 subjects with 1415 total daily outcome/feature pairs. Following Bair \emph{et al.} \cite{bair2006prediction}, the number of features is further reduced by selecting features with significant univariate correlations while controlling for a false discovery rate of 0.1.

\begin{figure}[H]
\begin{center}
\includegraphics[width=15cm]{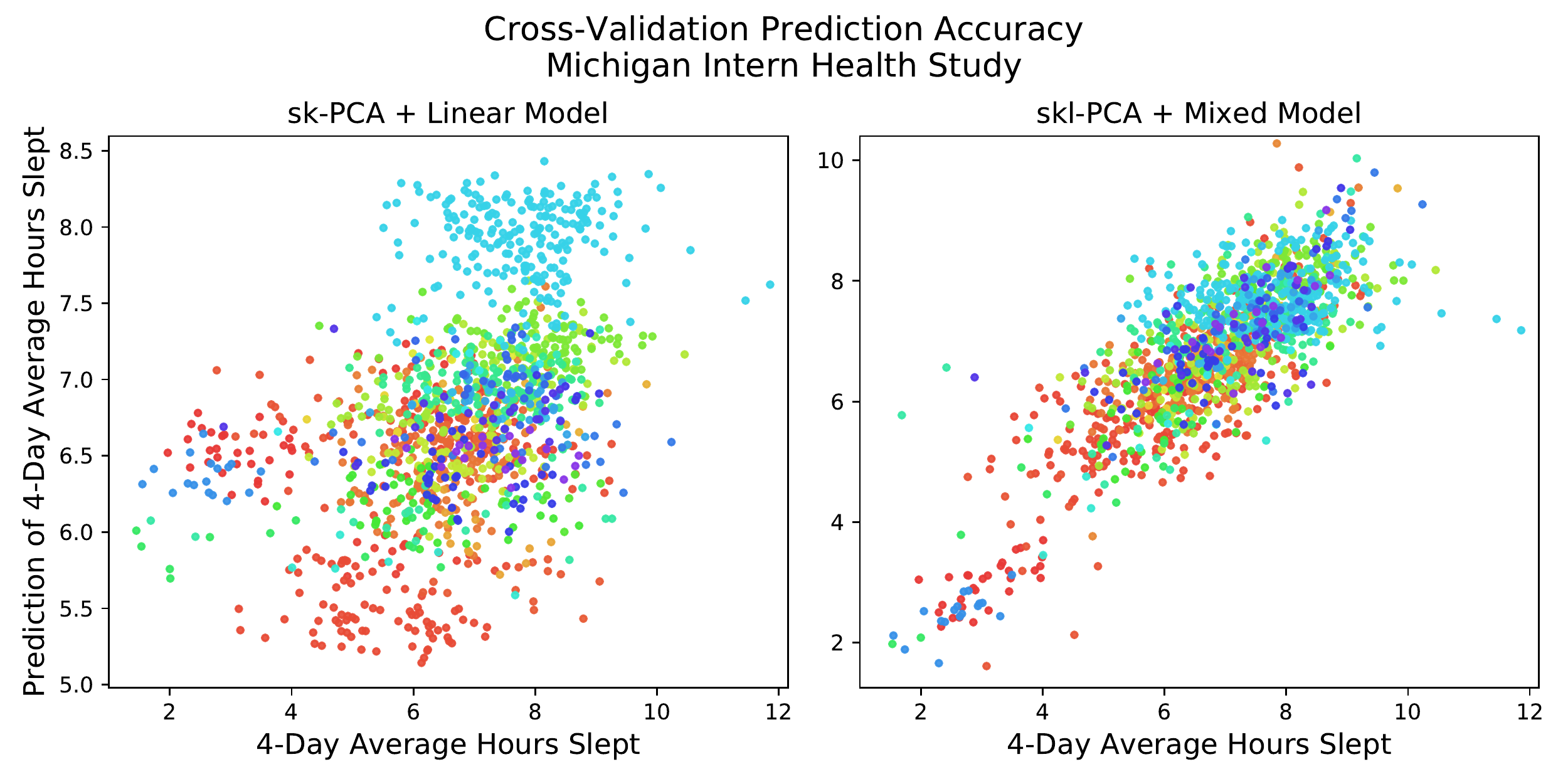}
\caption{The cross-validated correlation comparing two modeling approaches in the Michigan Intern Health Study. The left panel displays prediction results employing \emph{skPCA}, and the right panel displays results using \emph{sklPCA}. The $x$-axis depicts the 4-day average hours slept, and the $y$-axis represents the prediction. Each color represents a distinct subject. \emph{skPCA} shows less overall cross-validated correlation, as well as between-subject prediction differences that are broadly unrelated to the outcome. \emph{sklPCA} shows higher cross-validated correlations both within and between subjects.}
\label{figure:application_prediction_accuracy}
\end{center}
\end{figure}

\begin{figure}[H]
\begin{center}
\includegraphics[width=15cm]{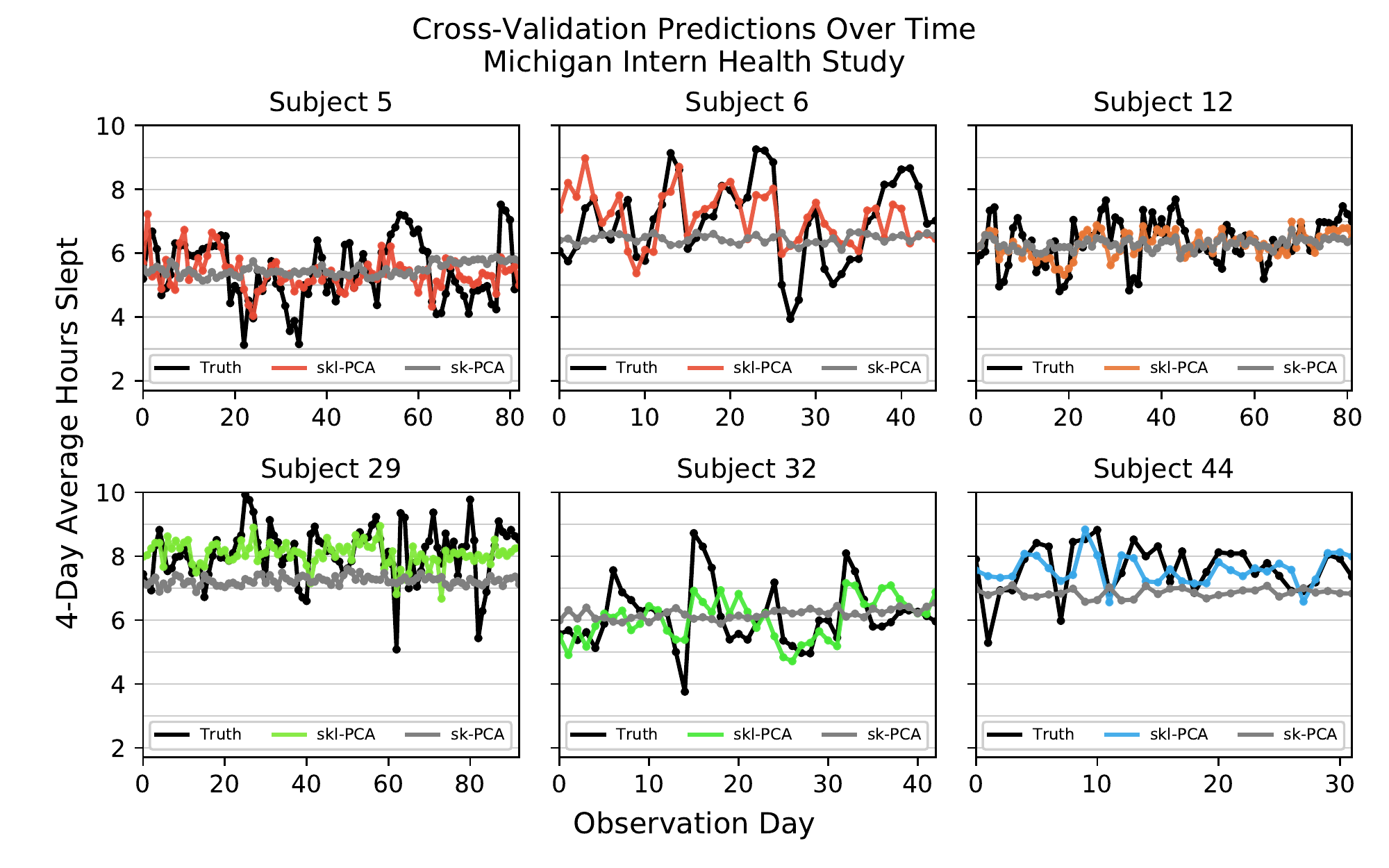}
\caption{Outcomes and cross-validated model predictions over time for a selection of subjects in the Michigan Intern Health Study. The six panels depict results for a single selected subject each. The $x$-axis shows days for which outcomes and biomarker data were both available for modeling, ordered in time. The $y$-axis shows the 4-day average of sleep for each observation. Black lines show measured sleep averages over time, gray lines show cross-validated predictions using \emph{skPCA}, and the variously colored lines per panel show cross-validated predictions using \emph{sklPCA}.}
\label{figure:application_predictions_over_time}
\end{center}
\end{figure}

After performing these preprocessing steps, we compare the two methods previously discussed in this section. Employing \emph{skPCA} yields a cross-validated correlation of 0.438 ($p$<1e-10). While using \emph{sklPCA}, a cross-validated correlation of 0.814 ($p$<1e-10) is achieved (see Figure \ref{figure:application_prediction_accuracy}). Individual predictions over time for a subset of subjects are shown in Figure \ref{figure:application_predictions_over_time}. Combined, these suggest that performing \emph{SKDR} while accounting for subject identity in longitudinal datasets can result in substantially improved model performance.

\vspace{-1cm}

\vspace{-1cm}

\section{Conclusion}

We have proposed a supervised kernel dimension reduction method in the longitudinal setting. In Section 2, we derived a novel decomposition for the \emph{HSIC}, and proposed \emph{sklPCA}, a method extending \emph{skPCA} that maximizes two constituent components of \emph{HSIC} separately. In Section 3, we showed that our method yields comparable or superior accuracy compared to \emph{skPCA} for a range of simulation settings. We also applied these approaches to the Michigan Intern Health Study, for which \emph{sklPCA} yielded a cross-validated correlation approximately twice that obtained using \emph{skPCA}.

This estimation procedure might be improved in several ways. This work extends the \emph{HSIC} estimator proposed by Gretton \emph{et al.,} \cite{gretton2005measuring}, yet more efficient estimators in both the computational and statistical sense have recently been proposed \cite{perez2017sensitivity}. Extending this approach for longitudinal data might be similarly efficient. Alternatively, the estimator for the \emph{HSIC} used here is based on $U$-statistics \cite{hoeffding1948class}, which have been extended to non-\emph{i.i.d.} data \cite{borovskikh1996u}, which might provide another route through which estimators of the \emph{HSIC} may be derived for clustered data. Our method transforms covariates by selecting generalized eigenvectors, but techniques for recasting variable transformation methods as feature selection methods have been developed \cite{masaeli2010transformation}, which might be adopted here. Finally, new kernels might be developed that account for within-subject clustering, such as an extension of the Gaussian mixture proposed by Weiss \emph{et al.} \cite{weiss1999segmentation}. We leave such investigations as future work.

\pagebreak
\section{Supplementary Materials}
In this supplement, we provide proofs for the theorems stated in the paper.
\subsection{Proof of Theorem 1}
\textbf{Theorem 1}: \emph{Let $X_1$, $X_2$, and $Y$ be random variables, $\mathcal{F}_1$, $\mathcal{F}_2$, and $\mathcal{G}$ be separable Hilbert spaces equipped with kernels $k_1: X_1\times X_1\rightarrow \mathbb{R}$, $k_2: X_2\times X_2\rightarrow \mathbb{R}$, and $\ell: Y\times Y\rightarrow \mathbb{R}$ such that $k_1(x_1, x_1') = \langle\varphi(x_1)$, $\varphi(x_1')\rangle_{\mathcal{F}_1}$, $k_2(x_2, x_2') = \langle\varphi(x_2), \varphi(x_2')\rangle_{\mathcal{F}_2}$, and $\ell(y, y') = \langle\psi(y), \psi(y')\rangle_{\mathcal{G}}$ for some explicit feature maps $\varphi_1: X_1 \rightarrow \mathcal{F}_1$, $\varphi_2: X_2\rightarrow \mathcal{F}_2$, and $\psi:Y\rightarrow \mathcal{G}$. Let $X_1\oplus X_2$ be the direct sum of these random variables, and $\mathcal{F}_1\oplus \mathcal{F}_2$ be the direct sum of these Hilbert spaces $\mathcal{F}_1$ and $\mathcal{F}_2$. Then:}

\begin{align}
\emph{HSIC}(\mathbb{P}_{X_1\oplus X_2, Y}, \mathcal{F}_1\oplus \mathcal{F}_2, \mathcal{G}) =  \emph{HSIC}(\mathbb{P}_{X_1, Y}, \mathcal{F}_1, \mathcal{G}) + \emph{HSIC}(\mathbb{P}_{X_2, Y}, \mathcal{F}_2, \mathcal{G}).
\end{align}

\emph{Proof}: The direct sum of two Hilbert spaces is also a Hilbert space, equipped with the inner product \newline
$\langle x_1\oplus x_2, x_1'\oplus x_2'\rangle_{\mathcal{F}_1\oplus\mathcal{F}_2}:=\langle x_1, x_1'\rangle_{\mathcal{F}_1}+\langle x_2', x_2'\rangle_{\mathcal{F}_2}$. The kernel for feature map $\varphi_1\oplus\varphi_2$ is therefore:

\begin{align}
k(x_1\oplus x_2, x_1'\oplus x_2') &= \langle\varphi_1(x_1)\oplus \varphi_2(x_2), \varphi_1(x_1')\oplus \varphi_2(x_2')\rangle_{\mathcal{F}_1\oplus \mathcal{F}_2} \\
&= \langle\varphi(x_1), \varphi(x_1')\rangle_{\mathcal{F}_1} + \langle\varphi_2(x_2), \varphi_2(x_2')\rangle_{\mathcal{F}_2} \\
&= k_1(x_1, x_1') + k_2(x_2, x_2')
\end{align}

Let $\mu_{\varphi(X)}:= \mathbb{E}_{X}\left[\varphi(x)\right]=\int_x\varphi(x)p_X(x)$. Using the linearity of expectations and the distributive properties for direct sums and tensor products, we obtain:

\begin{align}
&\mathbb{E}_{X_1\oplus X_2}[(\varphi_1(x_1)\oplus\varphi_2(x_2)-\mu_{\varphi_1(X_1)\oplus\varphi_2(X_2)})\otimes(\psi(y)-\mu_{\psi(Y)})] =\\
&\mathbb{E}_{X_1, Y}[(\varphi_1(x_1)-\mu_{\varphi_1(X_1)})\otimes(\psi(y)-\mu_{\psi(Y)})]\oplus \mathbb{E}_{X_2, Y}[(\varphi_2(x_2)-\mu_{\varphi_2(X_2)})\otimes(\psi(y)-\mu_{\psi(Y)})] =\\
&C_{x_1, y}\oplus C_{x_2, y}
\end{align}

for cross-covariance linear operators $C_{x_1, y}:\mathcal{G}\rightarrow \mathcal{F}_1$ and $C_{x_2, y}: \mathcal{G}\rightarrow \mathcal{F}_2$. The Hilbert-Schmidt norm of $C_{x_1, y}\oplus C_{x_2, y}$ is therefore:

\begin{align}
\|C_{x_1, y}\oplus C_{x_2, y}\|^2_{\text{HS}} &= \sum_{i, j, k} \langle C_{x_1, y}\oplus C_{x_2, y}(v_i\oplus w_j), u_k\rangle_{\mathcal{F}_1\oplus \mathcal{F}_2} \\
&= \sum_{i, k}\langle C_{x_1, y}v_i, u_k\rangle_{\mathcal{F}_1}+\sum_{j, k}\langle C_{x_2, y}w_j, u_k\rangle_{\mathcal{F}_2} \\
&= \emph{HSIC}(\mathbb{P}_{X_1, Y}, \mathcal{F}_1, \mathcal{G}) + \emph{HSIC}(\mathbb{P}_{X_2, Y}, \mathcal{F}_2, \mathcal{G})
\end{align}

\begin{flushright}
$\blacksquare$
\end{flushright}

\pagebreak
\subsection{Proof of Theorem 2}

\textbf{Theorem 2}: \emph{Let $X$, $Y$, and $Z$ be random variables, $\mathcal{F}$ and $\mathcal{G}$ be separable Hilbert spaces equipped with kernels $k: X\times X\rightarrow \mathbb{R}$ and $l: Y\times Y\rightarrow \mathbb{R}$ such that $k(x, x') = \langle\varphi(x), \varphi(x')\rangle_{\mathcal{F}}$ and $\ell(y, y') = \langle\psi(y), \psi(y')\rangle_{\mathcal{G}}$ for some explicit feature maps $\varphi:X\rightarrow \mathcal{F}$ and $\psi:Y\rightarrow \mathcal{G}$. Then:}

\begin{align}
\emph{HSIC}(\mathbb{P}_{X, Y}, \mathcal{F}, \mathcal{G}) &= \underbrace{\left\|\text{Cov}_Z(\mu_{\varphi(X\mid z)}, \mu_{\psi(Y\mid z)})\right\|_{\text{HS}}^2}_{\emph{HSIC}_{\text{fixed}}} + \underbrace{\mathbb{E}_Z\left\|\text{Cov}_{X, Y \mid z}(\varphi(x), \psi(y))\right\|_{\text{HS}}^2}_{\emph{HSIC}_{\text{random}}} + \zeta(\mathbb{P}_{X, Y, Z}, \mathcal{F}, \mathcal{G}). \label{theorem_statement_2}
\end{align}

\emph{Proof:} For ease of notation, define the following expectations: 

\begin{align}
\mu_{\varphi(X)}&:= \mathbb{E}_{X}\left[\varphi(x)\right]=\int_x\varphi(x)p_X(x) \\
\mu_{\varphi(X\mid z)}&:=\mathbb{E}_{X\mid z}\left[\varphi(x)\right]=\int_x\varphi(x)p_{X\mid z}(x\mid z)
\end{align}

Applying these to the cross-covariance operator for $\varphi$ and $\psi$, we obtain:

\begin{align}
\text{Cov}_{X, Y}(\varphi(x), \psi(y)) &= \mathbb{E}_{X, Y}\left[\varphi(x)\otimes \psi(y)\right]-\mu_{\varphi(X)}\otimes \mu_{\psi(Y)} \label{covariance}\\
&=\mathbb{E}_Z\left[\mathbb{E}_{X, Y \mid z}\left[\varphi(x) \otimes \psi(y)\right]\right] - \mathbb{E}_Z\left[\mu_{\varphi(X\mid z)}\right]\otimes\mathbb{E}_Z\left[\mu_{\psi(Y\mid z)}\right] \\
&=\mathbb{E}_Z\left[\text{Cov}_{X, Y \mid z}(\varphi(x), \psi(y)) + \mu_{\varphi(X\mid z)}\otimes \mu_{\psi(Y\mid z)}\right] - \mathbb{E}_Z\left[\mu_{\varphi(X\mid z)}\right]\otimes\mathbb{E}_Z\left[\mu_{\psi(Y\mid z)}\right] \\
&=\mathbb{E}_Z\left[\text{Cov}_{X, Y \mid z}(\varphi(x), \psi(y))\right] + \mathbb{E}_Z\left[\mu_{\varphi(X\mid z)}\otimes \mu_{\psi(Y\mid z)}\right] - \mathbb{E}_Z\left[\mu_{\varphi(X\mid z)}\right]\otimes\mathbb{E}_Z\left[\mu_{\psi(Y\mid z)}\right] \\
&=\mathbb{E}_Z\left[\text{Cov}_{X, Y \mid z}(\varphi(x), \psi(y))\right] + \text{Cov}_Z(\mu_{\varphi(X\mid z)}, \mu_{\psi(Y\mid z)}) \label{HSIC}
\end{align}

The norm of (\ref{covariance}), or $\emph{HSIC}(\mathbb{P}_{X, Y}, \mathcal{F}, \mathcal{G})$, decomposes into three terms:

\begin{align}
\left\|\text{Cov}_{X, Y}(\varphi(x), \psi(y))\right\|^2_{\text{HS}} &= \left\|\text{Cov}_Z(\mu_{\varphi(X\mid z)}, \mu_{\psi(Y\mid z)})\right\|_{\text{HS}}^2 + \left\|\mathbb{E}_Z\ \text{Cov}_{X, Y \mid z}(\varphi(x), \psi(y))\right\|_{\text{HS}}^2\nonumber \\
&\hspace{1cm} + 2\langle\mathbb{E}_Z[\text{Cov}_{X, Y\mid z}(\varphi(x), \psi(y))], \text{Cov}_Z(\mu_{\varphi(X\mid z)}, \mu_{\psi(Y\mid z)})\rangle_{\text{HS}} \label{final_equation}
\end{align}

The second term may be evaluated as follows:

\begin{align}
\left\|\mathbb{E}_Z\text{Cov}_{X, Y \mid z}(\varphi(x), \psi(y))\right\|_{\text{HS}}^2 &= \mathbb{E}_Z\left\|\text{Cov}_{X, Y \mid z}(\varphi(x), \psi(y))\right\|_{\text{HS}}^2 \nonumber \\
&\hspace{1cm} + 2 \sum_{z<z'}\left\langle \text{Cov}_{X, Y \mid z}(\varphi(x), \psi(y)), \text{Cov}_{X, Y \mid z'}(\varphi(x), \psi(y))\right\rangle_{\text{HS}}
\end{align}

Therefore, Equation (\ref{theorem_statement_2}) holds upon gathering the cross-covariance terms: 

\begin{align}
\zeta(\mathbb{P}_{X, Y, Z}, \mathcal{F}, \mathcal{G}) := &-2\langle\mathbb{E}_Z[\text{Cov}_{X, Y\mid z}(\varphi(x), \psi(y))], \text{Cov}_Z(\mu_{\varphi(X\mid z)}, \mu_{\psi(Y\mid z)})\rangle_{\text{HS}} \nonumber \\
&+2 \sum_{z<z'}\left\langle \text{Cov}_{X, Y \mid z}(\varphi(x), \psi(y)), \text{Cov}_{X, Y \mid z'}(\varphi(x), \psi(y))\right\rangle_{\text{HS}}
\end{align}

\begin{flushright}
$\blacksquare$
\end{flushright}

\pagebreak
\subsection{Proof of Theorem 3}

\textbf{Theorem 3}: \emph{Under regulatory conditions, then $\widehat{\text{HSIC}}_{\emph{fixed}} := (m-1)^{-2}\emph{tr}(\overline{\mathbf{K}}_m\mathbf{H}_m\overline{\mathbf{L}}_m\mathbf{H}_m)$ is a consistent estimator of $\text{HSIC}_{\emph{fixed}}$ with bias of order $\mathcal{O}(m^{-1})$ and convergence rate $\mathcal{O}(m^{-1/2})$.}

\emph{Proof:} $\emph{HSIC}_{\text{fixed}}$ may be written in terms of conditional expectations:

\begin{align}
\emph{HSIC}_{\text{fixed}} &= \left\|\text{Cov}_Z(\mu_{\varphi(X\mid z)}, \mu_{\psi(Y\mid z)})\right\|_{\text{HS}}^2 \\
&= \mathbb{E}_{Z,Z'}\left[\mathbb{E}_{X\mid z, X'\mid z'}\left[k(x,x')\right]\mathbb{E}_{Y\mid z, Y'\mid z'}\left[\ell(y,y')\right]\right] \nonumber \\
&\hspace{1cm}-2\mathbb{E}_{Z}\left[\mathbb{E}_{X\mid z, X'}\left[k(x,x')\right]\mathbb{E}_{Y\mid z, Y'}\left[\ell(y,y')\right]\right] \nonumber \\
&\hspace{1cm}+\mathbb{E}_{X, X'}[k(x,x')]\mathbb{E}_{Y, Y'}[\ell(y,y')] \\
&=\mathbb{E}_{Z,Z'}\left[\mathbb{E}_{X\mid z, X'\mid z'}\left[k(x,x')\right]\mathbb{E}_{Y\mid z, Y'\mid z'}\left[\ell(y,y')\right]\right] \nonumber \\
&\hspace{1cm}-2\mathbb{E}_{Z}\left[\mathbb{E}_{Z'}\left[\mathbb{E}_{X\mid z, X'|z'}\left[k(x,x')\right]\right]\mathbb{E}_{Z'}\left[\mathbb{E}_{Y\mid z, Y'|z'}\left[\ell(y,y')\right]\right]\right] \nonumber \\
&\hspace{1cm}+\mathbb{E}_{Z, Z'}\left[\mathbb{E}_{X|z, X'|z'}[k(x,x')]\right]\mathbb{E}_{Z, Z'}\left[\mathbb{E}_{Y|z, Y'|z'}[\ell(y,y')]\right]
\end{align}

$\widehat{\emph{HSIC}}_{\text{fixed}}$ may be decomposed as follows:

\begin{align}
\widehat{\emph{HSIC}}_{\text{fixed}} &= (m-1)^{-2}\text{tr}\left(\overline{\mathbf{K}}_m\mathbf{H}_m\overline{\mathbf{L}}_m\mathbf{H}_m\right) \\
&= (m-1)^{-2}\text{tr}(\overline{\mathbf{K}}_m\overline{\mathbf{L}}_m) + (m-1)^{-3}\mathbf{1}^\top\overline{\mathbf{K}}_m\overline{\mathbf{L}}_m\mathbf{1} + (m-1)^{-4}\mathbf{1}^\top\overline{\mathbf{K}}_m\mathbf{1}\mathbf{1}^\top\overline{\mathbf{L}}_m\mathbf{1} \\
&= \frac{1}{(m-1)^2}\sum_{i,i'=1}^m\overline{\mathbf{K}}_{i,i'}\overline{\mathbf{L}}_{i,i'} \nonumber \\
&\hspace{1cm}+ \frac{1}{m}\sum_{i=1}^m \frac{1}{(m-1)^2}\sum_{i',i''=1}^m \overline{\mathbf{K}}_{i',i}\overline{\mathbf{L}}_{i,i''} \nonumber \\
&\hspace{1cm}+ \left(\frac{1}{m(m-1)}\sum_{i,i'=1}^m\overline{\mathbf{K}}_{i,i'}\right)\left(\frac{1}{m(m-1)}\sum_{i,i'=1}^m\overline{\mathbf{L}}_{i,i'}\right)
\end{align}

Therefore, $\widehat{\emph{HSIC}}_{\text{fixed}}$ is the sum of $U$-estimators in terms of $\overline{\mathbf{K}}_{i,i'}$ and $\overline{\mathbf{L}}_{i,i'}$. If these terms are bounded, then by Theorems 1 and 3 in \cite{gretton2005measuring}, it follows immediately that $\widehat{\emph{HSIC}}_{\text{fixed}}$ has bias $\mathcal{O}(m^{-1})$ and converges to $\emph{HSIC}_{\text{fixed}}$ with rate $\mathcal{O}(m^{-1/2})$.

As $\forall i\ (n_i\rightarrow \infty)$, the regulatory conditions can be relaxed: only $\overline{k}_{i,i'}$ and $\overline{\ell}_{i,i'}$ must be bounded. To show this, we prove $\overline{\mathbf{K}}_{i,i'}\overset{p}{\longrightarrow} \overline{k}_{i,i'}$ and $\overline{\mathbf{L}}_{i,i'}\overset{p}{\longrightarrow} \overline{\ell}_{i,i'}$. Assume $k$ is bounded by $B_k$. The (unbiased) $U$-estimator for $\overline{k}_{i,i'}$ is $\overline{\mathbf{K}}^*_{i,i'} := \overline{\mathbf{K}}_{i,i'} - (n_i-1)^{-2}\text{tr}\left(\mathbf{K}_{i}\right)$, therefore $\overline{\mathbf{K}}_{i,i'}$ exhibits a bias of $\mathcal{O}(n_i^{-1})$. Furthermore, by \cite{hoeffding1963probability}:

\begin{align}
\mathbb{P}(\overline{\mathbf{K}}^*_{i,i'}-\overline{k}_{i,i'} \geq t)\leq \text{exp}\left(-n_it^2/B_k^2\right):=\delta
\end{align}

Solving for $t$, we obtain $t=B_k\sqrt{\frac{-\log(\delta)}{n_i}}$. $\overline{\mathbf{K}}^*_{i,i'}$ can be bounded below with error probability $\delta$ in similar fashion, thus $\overline{\mathbf{K}}_{i,i'}^*\overset{p}{\longrightarrow} k_{i,i'}$ with rate $\mathcal{O}(n_i^{-1/2})$, as does $\overline{\mathbf{K}}_{i,i'}$ by Slutsky's Theorem. Similarly, if $\ell$ is bounded, the bias and rate of convergence of $\overline{\mathbf{L}}_{i,i'}$ is $\mathcal{O}(n_i^{-1})$ and $\mathcal{O}(n_i^{-1/2})$ respectively, completing the proof.

\begin{flushright}
$\blacksquare$
\end{flushright}

\pagebreak
 \newcommand{\etalchar}[1]{$^{#1}$}

\end{document}